\documentstyle[12pt]{article}
\def\mpl{{m_{\rm Pl}}}

\def\la{\mathrel{\mathpalette\fun <}}

\def\fun#1#2{\lower3.6pt\vbox{\baselineskip0pt\lineskip.9pt
  \ialign{$\mathsurround=0pt#1\hfil##\hfil$\crcr#2\crcr\sim\crcr}}}

\begin{document}
\pagestyle{empty}
\begin{center}
\bigskip

%\rightline{FERMILAB--Pub--98/***-A}
%\rightline{astro-ph/0212281}
%\rightline{submitted to {\it }}

\vspace{.2in}
{\Large \bf The New Cosmology:\\
\medskip
Mid-term Report Card for Inflation}
\bigskip

\vspace{.2in}
Michael S. Turner\\

\vspace{.2in}
{\it Departments of Astronomy \& Astrophysics and of Physics,\\
Center for Cosmological Physics, and \\
Enrico Fermi Institute, The University of Chicago, Chicago, IL~~60637-1433}\\

\vspace{0.1in}
{\it NASA/Fermilab Astrophysics Center\\
Fermi National Accelerator Laboratory, Batavia, IL~~60510-0500}\\

\end{center}

\vspace{.3in}
\centerline{\bf ABSTRACT}
\bigskip

Inflation has been the driving idea in
cosmology for two decades and is a pillar
of the New Cosmology.  The inflationary paradigm
has now passed its first
round of significant tests, with two of its three basics
predictions confirmed at about the 10\% level (spatially
flat Universe and density perturbations produced
from quantum fluctuations with $|n-1|\sim {\cal O}(0.1)$).
{\em The Inflationary Paradigm has some of the truth.}  Over the next decade
the precision of these tests, most of which involve
measurements of CMB anisotropy and polarization, will improve 30 fold
or more(!), testing inflation more sharply and possibly
elucidating the underlying cause.  Especially important
in this regard is detecting the inflation-produced
gravitational waves, either directly or through their
CMB polarization signature.  While inflation
has by no means been verified, its successes have raised the bar
for competitor theories:  Any alternative must feature the two hallmarks
of inflation: superluminal expansion and entropy production.

\newpage
\pagestyle{plain}
\setcounter{page}{1}
\newpage

\section{Introduction}

Today, cosmology has a comprehensive and self-consistent
mathematical model that accounts for all the observed
features of the Universe.  However, unlike its predecessor,
the hot big-bang model or standard cosmology, there is no
standard name.  For now, I will refer to it
as the New Cosmology \cite{NewCosmology}.

The New Cosmology incorporates the hot big-bang model
(every successor theory eats its predecessor whole!),
as well as an early inflationary epoch and the
present stage of accelerated expansion.  The New
Cosmology includes a full accounting of the shape and composition
of the Universe
today:  spatially flat, with the critical density distributed
as follows:  baryons ($4\pm 1\%$), nonbaryonic dark matter
($29\pm 4\%$) and dark energy ($67\pm 6\%$) \cite{turner02}.

While we can now say that massive neutrinos account
for between $0.1\%$ and $5\%$ \cite{omega_neutrino}
of the nonbaryonic dark matter -- comparable to the $0.5\%$
contributed by stars -- most of the dark matter is thought
to be slowly moving elementary particles (cold dark
matter), with the leading candidates being the
axion and neutralino \cite{dmreview}.

Dark energy is the name I use for the mysterious
``energy stuff'' that dominates the mass-energy budget and whose
large negative pressure ($p<-\rho/2$) is responsible
for the accelerated expansion \cite{turner_de}.  Dark energy could
be as ``mundane'' as the the energy of the quantum vacuum,
or something so exotic that it has thus far eluded
the minds of the most creative theorists \cite{carroll_review}.
Dark energy is truly one of the great mysteries in all of science.

The standard cosmology can properly claim to
give a reliable and tested description of the Universe
from a fraction of a second onward. While the New Cosmology
cannot yet make a similar claim for extending our understanding
back to an early, inflationary epoch, inflation is nonetheless a pillar
of the New Cosmology, and, as I will describe, precision CMB
observations are beginning to test its basic
predictions (so far, so good).  This is a remarkable development that
belies the prediction made by some astronomers (and the
fear of many inflationary theorists) that inflation
would never be tested.

\section{The Inflationary Paradigm}

Few ideas in theoretical physics have had as much impact as inflation.  Introduced by
Alan Guth in a 1980 paper \cite{guth} that explained the cosmological virtues
of exponential expansion as well as why his version of it (based
upon a symmetry-breaking phase transition) did {\em not} work(!), inflation
has been the driving idea in cosmology since.

The virtues of inflation trace to its ability to lessen the dependence of the
present state of the Universe upon initial conditions (though quantifying
how successful it is at achieving this is difficult \cite{nohair}).
In addition to the predictions
discussed below, it explains the high degree of homogeneity and isotropy
observed in our Hubble volume, the heat of the big bang, and the absence of
superheavy magnetic monopoles.

In a flurry of activity during the early 1980s the basic inflation
paradigm \cite{KT} was worked out \cite{slowroll,scalar,tensor,reheat}.
In brief, essentially all models of inflation can be described in terms
of the classical evolution of a single scalar field (dubbed the inflaton)
initially displaced from the minimum of its potential [$V(\phi )$]:
\begin{eqnarray}
\ddot\phi + 3H\dot\phi + V^\prime & = & 0 \\
N \equiv  \int \, Hdt & = & {8\pi \over \mpl^2}\int\,{V(\phi )d\phi \over V^\prime (\phi )}
\end{eqnarray}
where $N$ is the number of e-folds of inflation, prime denotes
$d/d\phi$ and dot denotes $d/dt$.  Models of inflation
differ only in the form of their scalar potential and their motivations.

Inflation occurs while the inflaton is slowly rolling:
during the ``slow roll,'' the {\em nearly} constant potential energy density
associated with the inflaton drives a nearly exponential expansion
(``superluminal expansion'').  At
the end of the slow roll, the inflaton is left oscillating about its potential
energy minimum; these oscillations correspond to a condensate
of zero-momentum $\phi$ particles.  Their decay (by coherent and/or
incoherent processes) produces lighter particles which thermalize;
this ``reheats'' the Universe and exponentially increases its entropy
(per comoving volume).  (Note:  The Universe need not be hot before inflation.)  
The standard hot big bang phase commences thereafter, with the 
tremendous entropy release accounting its heat.

Quantum fluctuations in $\phi$ ($\Delta \phi \sim H/2\pi$) give rise to
energy density fluctuations ($\delta \rho = \Delta \phi V^\prime$),
which ultimately result in inflation's signature adiabatic density perturbations.
They are of astrophysical interest because of their physical size is stretched
from the subatomic to the astrophysical during the period of exponential
expansion.  Similarly, quantum fluctuations in the metric itself ($h\sim
\delta g \sim H/m_{\rm Pl}$) give
rise to a spectrum of gravitational waves with wavelengths of astrophysical
interest (fluctuations in other light scalar fields can result in
particle production or isocurvature perturbations).

Inflation is a paradigm and not a model because there is no
agreed upon identity for the inflaton field.
Many models exist, with the inflaton
playing a variety of roles, from inducing electroweak symmetry breaking
to breaking supersymmetry to the compactification of extra dimension(s),
and the energy scale of inflation ranging from
$1\,$TeV to $10^{16}\,$GeV \cite{LL}.  {\em While there is no standard model,
each model makes its own set of precise predictions.}

That being said, the inflationary paradigm does make a set of
generic predictions that can be used to test -- and even falsify -- it.
The basic predictions of inflation are:
\begin{itemize}
\item Spatially flat Universe
\item Not quite scale-invariant, almost power-law
spectrum of Gaussian, adiabatic density perturbations
\item Not quite scale-invariant, almost power-law
spectrum of gravitational waves
\end{itemize}
The first two of these predictions are now being tested by
precision measurements of cosmic microwave background (CMB)
anisotropy on sub-degree angular scales, with
early results consistent with inflation.  The third prediction,
which may hold the key to definitively testing inflation and shedding light
upon the underlying cause, is inspiring a new generation of
very challenging experiments.

\section{Specific models make specific predictions \cite{10things}}

While all models of inflation are based upon speculative physics that
goes beyond (usually well beyond) the standard model of particle
physics, each model makes predictions that sharpen the
basic predictions of the paradigm.  The reason for this is simple:
The physics is speculative but the rules are well defined --
the semi-classical evolution of a scalar field.

While models have been constructed where two or more fields
evolve during inflation, or where the kinetic term for the
inflaton is not canonical, I will discuss the predictions
for single-field inflation, assuming that the kinetic term
is canonical.  More complicated models also make definite predictions,
though the relationship of observable quantities to the potential(s)
can be different.

The prediction of a flat Universe is not tied to the form
of the potential; only that inflation lasts sufficiently
long to explain the homogeneity and isotropy.  Generally
speaking, the duration of inflation far exceeds that needed
to produce a flat Universe (although models of inflation have
been tuned to give $\Omega_0$ less than 1).  The inflationary
prediction of a flat Universe corresponds to density parameter
$\Omega_0 = 1$, where $\Omega_0$ is the ratio of the total
matter/energy density to the critical density.

The other two predictions of inflation involve
the metric perturbations that arise from the quantum fluctuations
associated with deSitter space:
adiabatic density (or scalar) perturbations from fluctuations
in the inflaton potential energy and gravity
waves (tensor perturbations) from fluctuations in the
metric itself.  Their amplitude and variation
with scale do depend upon the properties of the scalar-field potential, and
this is the basis for the belief that observations may someday pin
down the underlying model of inflation \cite{recon}

Both the scalar and tensor perturbations have an
approximately -- but not exactly -- scale-invariant spectrum.
This fact fundamentally
traces to the approximate deSitter space associated with the
inflationary phase.  In physical terms, that means
that the dimensionless strain amplitude of gravity waves when they
re-enter the horizon after inflation is almost independent of scale:
$h_{\rm HOR}\sim
H/\mpl \sim V^{1/2}/\mpl^2$.  For density perturbations, it is the
amplitude of the density perturbation at horizon crossing that is almost
independent of scale: $(\delta\rho /\rho )_{\rm HOR} \sim H^2/\dot\phi
\sim V^{3/2}/\mpl^3 V^\prime$.  Because the post-horizon-crossing
evolution of both density perturbations and gravity waves
depends upon the time elapsed since horizon crossing -- which
is longer for longer wavelength perturbations -- the spectra of
density perturbations and gravity waves is not scale invariant
today.  The Fourier components of both scalar and tensor
perturbations are approximately power-law in wavenumber $k$ of
the fluctuations; at horizon crossing:
\begin{eqnarray*}
(\delta \rho /\rho)_{\rm HOR} & \ \propto \ & k^{(n-1)/2} \\
h_{\rm HOR} & \ \propto \ & k^{n_T/2}
\end{eqnarray*}
where $n$ ($n_T$) are the power-law indices for scalar
(tensor) perturbations.

Both density perturbations and gravity waves lead to fluctuations
in the temperature and polarization of the CMB across the sky.
Predictions for observable quantities can be expressed in terms of the inflationary
potential.  For instance, the scalar ($S$) and the tensor
contributions ($T$) to the CMB quadrupole anisotropy are
\begin{eqnarray}
S  & \equiv &  {5C_2^S \over 4\pi} \simeq
         2.9\,{V/\mpl^4 \over (\mpl V^\prime /V)^2} \\
T  & \equiv & {5C_2^T \over 4\pi} \simeq
        0.56 (V/\mpl^4)
\end{eqnarray}
where $C_2^S$ and $C_2^T$ are the contribution of scalar and tensor
perturbations to the variance of the $l=2$ multipole amplitude
($\langle |a_{2m}|^2\rangle = C_2^S+C_2^T$) and
$V$ is the value of the inflationary potential when the scale
$k=H_0$ (present horizon scale) crossed the Hubble radius during inflation.
The numerical coefficients in these expressions depend slightly
upon the composition of the Universe; the numbers shown are for
$\Omega_M = 0.35$ and $\Omega_\Lambda = 0.65$ \cite{TW}.

The power-law indices
that characterize the scalar and gravity-wave spectra can be expressed
in terms of the inflationary potential and its derivatives:
\begin{eqnarray}
n -1 & = & -{1\over 8\pi}\left({\mpl V^\prime \over V} \right)^2
    + {\mpl \over 4\pi}\left( {\mpl V^{\prime}\over V} \right)^\prime \\
n_T & = & -{1\over 8\pi} \left( {\mpl V^\prime \over V} \right)^2
\end{eqnarray}
For typical inflationary potentials the deviations
from scale invariance are expected to be of order 10\%:  $|n-1|\sim
{\cal O}(0.1)$ and $n_T \sim -{\cal O}(0.1)$ \cite{ht}.

Note that the ratio between the gravity-wave contribution to the CMB
quadrupole anisotropy and the density-perturbation contribution to the
CMB quadrupole anisotropy provides a consistency test if the tensor
spectral index can be measured:
\begin{equation}
T/S = -5n_T
\end{equation}
for $\Omega_M = 0.35$ and $\Omega_\Lambda = 0.65$ \cite{TW}.

The fluctuation spectra are not exact power laws (except in the case
of power-law inflation); variations in the power-law indices with $k$
may be expressed in terms of higher derivatives of $V(\phi )$ \cite{kt}.
\begin{eqnarray}
{dn\over d\ln k}& = & -{\mpl^2\over 8\pi}\left({V^\prime \over V}\right)
        {dn \over d\phi}\nonumber\\
                & = &
-{1\over 32\pi^2}\left({{m_{\rm Pl}}^3V_*^{\prime\prime\prime}\over V_*}\right)
        \left({{m_{\rm Pl}} V_*^\prime\over V_*}\right) \nonumber\\
     & & \hspace*{1cm} + {1\over 8\pi^2}
        \left({{m_{\rm Pl}}^2V_*^{\prime\prime}\over V_*}\right)
        \left({{m_{\rm Pl}} V_*^\prime\over V_*}\right)^2
        - {3\over 32\pi^2}\left({m_{\rm Pl}} {V_*^\prime\over V_*}\right)^4  \\
{dn_T\over d\ln k} & = & -{\mpl^2\over 8\pi}\left({V^\prime \over V}\right)
		{dn_T\over d\phi}\nonumber\\
& = & {1\over 32\pi^2}\left({{m_{\rm Pl}}^2V^{\prime\prime}\over V}
\right) \left( {{m_{\rm Pl}} V^\prime\over V}\right)^2  - {1\over 32\pi^2}
\left({{m_{\rm Pl}} V^\prime\over V}\right)^4
\end{eqnarray}

Finally, once the form of the density perturbations and the composition of
the Universe are specified, the initial conditions for the formation of
structure in the Universe are set.  Inflation specifies the form of the
density perturbations and the composition of the Universe is known -- 29\% cold
dark matter (with a dash of it in massive neutrinos),
4\% baryons and 66\% dark energy.  This means that the
large (and growing) body of observational data that probes the formation of structure
in the Universe provides an additional test of the inflationary paradigm.
In particular, they have significant potential to determine $n$ and
test the Gaussian nature of the underlying density perturbations.

\section{Mid-term report card and future expectations}

The three basic predictions of inflation blossom into a series of
9 testable consequences, most of which can also probe the underlying
inflationary model.

The cosmic microwave background will play the leading in testing these
predictions because its anisotropy and polarization are such a clean
probe of the gravity waves and density perturbations produced by inflation.
The predictions of inflationary models can be stated in terms of the
predicted variances of the multipole amplitudes that characterize
CMB anisotropy and polarization \cite{hudodelsonreview}; because
there are only $2\ell + 1$ multipole amplitudes for multipole $\ell$,
even an ideal experiment is limited by sample size (often referred
to as cosmic variance) in estimating the true variance.

The discovery of CMB anisotropy on angular scales of order 10 degrees
by the COBE Differential Microwave Radiometer in 1992 \cite{DMR} opened the door for testing
inflation.  The measurement of CMB anisotropy on sub-degree angular
scales by balloon-borne and ground-based experiments allowed the serious
testing to begin (most of the features and probative power lies in
the anisotropy on sub-degree scales) \cite{hudodelsonreview}.  Presently,
it is the results of the BOOMERanG, DASI, CBI, Maxima, CAT, and Archeops experiments
that define the state-of-the-art in our knowledge of the CMB angular
power spectrum.  Soon, the MAP all-sky satellite-based experiment will report
its first results and really clean up the anisotropy power spectrum
out to $\ell \sim 900$. The ESA/NASA Planck satellite is scheduled for launch in 2007; it should
provide the definitive power spectrum out to $\ell \sim 3000$ as well as
significant results on polarization.

The following are the nine predictions, the present status report and prospects
for the future.  A Table summary is given at the end of this section.

\medskip

\noindent{\bf 1. Spatial flatness}

\noindent This prediction is the most straightforward; it simply implies
$\Omega_0 = 1.0 \pm 0.00001$.  The `$\pm 0.00001$' arises because
fluctuations on the current Hubble scale will led to the knowable
part of the Universe appearing slightly open (underdensity) or
slightly closed (overdensity).

\noindent{\em Now:}  Measurements of the position of the
first acoustic peak in the CMB power spectrum indicate that
$\Omega_0 = 1.03\pm 0.03$, consistent with spatial flatness \cite{sieversetal}.
Further, a direct accounting of the amount of matter and energy
leads to an independent, though less precise, determination
of the total mass/energy that is reassuringly consistent with spatial
flatness, $\Omega_0 = 1\pm 0.25$.

\noindent{\em Future:}  The precision testing of this prediction lies with MAP, Planck
and other future CMB experiments that will probe the smallest angular
scales and fix the positions of the acoustic peaks with high precision.
Since the positions of the acoustic peaks also depend upon the
composition of the Universe, information from large-scale structure
measurements that constrains the matter density is also critical.
The ultimate precision to which $\Omega_0$ can be probed will
likely be in the range $\sigma_{\Omega_0} \sim 0.005 - 0.001$
and will depend upon how well the composition
can be fixed by other independent methods \cite{eht,hu}.

\bigskip
\noindent{\bf 2. Density Perturbations from Quantum Fluctuations}

\noindent The most striking prediction of inflation may be that
the density perturbations that seeded structure on scales of
millions of light years and larger arose
from quantum fluctuations on subatomic scales;  if this is true,
the CMB is a picture of quantum noise(!).  
This prediction breaks down into 5 separate testable consequences.

\medskip
\noindent{\bf a.  Acoustic peaks:}  Since the density perturbations are
impressed at very early times ($\ll {\rm sec}$), by the time of last scattering
($t\sim 400,000\,$yrs) all perturbations
are purely ``growing mode,'' leading to a synchronizing of the perturbations
on different scales.  Some modes were caught at maximum compression or
rarefaction, leading to the prediction of a series of ``acoustic peaks'' in
the angular (multipole) power spectrum \cite{hudodelsonreview}.

\noindent{\em Now:}  Current CMB experiments have probed the angular power
spectrum out to $\ell \sim 2000$;
at least three and perhaps as many five peaks have been resolved \cite{sieversetal}.
There is no question that the acoustic peaks associated adiabatic perturbations
have been seen.

\noindent{\em Future:}  Planck and other future CMB experiments that probe the smallest
angular scales should resolve six or more acoustic peaks (the damping of anisotropy on
very small angular scales due to the finite thickness of the last scattering
surface exponentially diminishes the amplitude of successive peaks).  These experiments
will also be able to separate out any small admixture of isocurvature perturbations
(which could arise during inflation or later on).

\medskip
\noindent{\bf b. Gaussianity:}  The inflation-produced density
perturbations arise from quantum fluctuations in a very weakly
coupled (essentially free) scalar field
and hence should be Gaussian to a high degree of precision.  The CMB has
the greatest power to test this prediction since it
probes the density perturbations when they were linear (nonGaussianity automatically
develops when gravity drives the amplitude of the perturbations into the nonlinear
regime).

\noindent{\em Now:}  There is no evidence for nonGaussianity.

\noindent{\em Future:}  The all-sky CMB mapping experiments (MAP and Planck) which
are designed to control systematics have the greatest the potential to test this
prediction.  Important to quantifying how well the Gaussianity prediction is
faring is the construction of a realistic model with nonGaussianity to compare
with.

\medskip
\noindent{\bf c.  Almost scale-invariant spectrum:}  Inflation predicts an almost,
but not quite scale invariant spectrum \cite{slowroll}.  Typically, the deviations from
scale invariance are of order 10\%:  $|n-1|\sim {\cal O}(0.1)$, with indeterminate sign \cite{ht}.
This is not only a key test of inflation, but a window to the underlying physics.

\noindent{\em Now:}  Measurements of CMB anisotropy at small angular scales
can probe this prediction most sharply.  Current observations are beginning
to significantly constrain $n$:  $n= 1.05\pm 0.09$ \cite{sieversetal}, consistent with inflation
but not precise enough to test the inflationary prediction sharply.

\noindent{\em Future:}  The Planck Mission should be able to determine $n$ to
a precision of $\pm 0.008$ \cite{eht}.  An ideal CMB experiment could
achieve a precision of almost ten times better \cite{hu};
whether or not a future ground-based and space-based experiment with such
capability is carried out remains to be seen.

\medskip
\noindent{\bf d.  Almost power-law:}  Inflation predicts that the power-law index varies
with scale, with $dn/d\ln k \sim \pm 10^{-3}$ for many models and ten times
larger for some \cite{kt}.

\noindent{\em Now:}  Current CMB measurements, $dn/d\ln k  = -0.02\pm 0.04$ \cite{run},
are consistent with the inflationary prediction, but lack the precision
to measure a variation.

\noindent{\em Future:}  It is likely that the CMB offers is the most powerful
probe, with Planck projected to do a factor of ten better than the current
limit \cite{eht}; an ideal CMB experiment might reach the
expected level of variation in $n$.

\medskip
\noindent{\bf e.  CDM scenario for structure formation:}  Nearly scale-invariant,
Gaussian adiabatic density perturbations is one of the two pillars of the
highly successful CDM paradigm for structure formation (the other being slowly
moving, weakly interacting dark matter particles).  As such, tests of the CDM
paradigm are tests of inflation.  In particular, the study of large-scale
can constrain $n$ and the abundance of rare objects, such as clusters, can
be used to test the Gaussianity prediction.

\bigskip
\noindent{\bf 3. Gravity Waves from Quantum Fluctuations}

\medskip
\noindent{\bf a.  Amplitude:}  The amplitude of the gravity waves
is directly proportional to the energy scale of inflation and does not depend upon
the shape of the potential.  Unfortunately, theory gives very little advice about
the amplitude, and unlike density perturbations which are de rigueur
(to seed structure formation) the Universe can live just fine without
gravity waves.  Huterer and I have explored what can be said about $T/S$
without regard to specific potentials by reformulating the equations
of motion for inflation.  We concluded that if $n>0.9$ and the potential
has no unnatural features (the key here is the definition of unnatural),
$T/S$ must exceed $10^{-3}$ \cite{ht2}.

\noindent{\em Now:}  There is no evidence for inflation-produced gravity waves;
current limits from the CMB imply $T/S < {\cal O}(1)$ \cite{kinneyetal}.

\noindent{\em Future:}  There are two basic ways to get at the
gravity waves.  The first is to use their signature in CMB anisotropy and polarization.
Unlike the series of acoustic peaks associated with density perturbations that
extend out to $\ell \sim 3000$, gravity waves produce a rather featureless angular
power spectrum that dies off at around $\ell \sim 100$.  Because the precision of even
a perfect CMB experiment is limited by  sampling variance,
CMB anisotropy alone can separate the gravity-wave
signature only if $T/S > 0.1$ \cite{knoxturner}.  However, the polarization signature
is not so limited:  gravity waves excite a mode of polarization (B-mode or curl)
that density perturbations cannot \cite{bmode}.  The only limitations to detecting gravity
waves through their polarization signature are sensitivity and foregrounds.
It appears that a dedicated polarization experiment might be able to achieve
a sensitivity to $T/S$ as good as $10^{-3}$ \cite{KnoxLimit} (recall, that if $n>0.9$
and the potential is ``natural'', $T/S$ is expected to be this large).
By comparison, Planck is expected to achieve $T/S \sim 0.02$ \cite{eht}.

The CMB is only sensitive to the longest wavelength gravity waves, $\lambda \sim
10^{26}\,$cm to $10^{28}$\,cm; however, the spectrum extends to wavelengths as
short as 1\,km.  Direct detection of inflation-produced gravity waves will
be very challenging as $\Omega_{\rm GW}$ is at most $10^{-15}$ at the mHz-kHz frequencies
where the planned detectors will operate: the projected
sensitivities of LIGO and LISA to a stochastic background of gravitational
waves are $10^{-10}$ and $10^{-12}$ respectively \cite{turnerGW}, far
short of what is expected.

\medskip
\noindent{\bf b. Spectrum:}  The predicted spectral index for one-field models
is related to the amplitude, $n_T = - {1\over 5}{T\over S}$, which
allows a consistency test if $n_T$ can be measured.  The combination of direct
and CMB detections could measure $n_T$ accurately:  owing to the
long lever-arm between the Hz frequencies of gravity-wave detectors and the
very long wavelength of gravity waves that affect the CMB, a factor of 2 precision
in each measurement would result in a few percent measurement of $n_T$.

\medskip
\noindent{\bf c. Consistency:}  For single-field inflation models $n_T =
-{1\over 5}{T\over S}$, as noted above.
\bigskip

\bigskip
\begin{table}
\caption{\bf TESTING INFLATION:  YEAR 2002 SUMMARY}
\vspace{2pc}
%%\begin{center}
%%\label{tab:report_card}
%%\scriptsize
\begin{tabular}{|l|c|c|}\hline
{Prediction} & {Mid-Term} &  {Expectation}  \\
\hline
\hline
%%\rule{0pt}{10pt}%
1. Flatness & & \\
{\ \ \ }a. $\Omega_0 = 1$ & $\Omega_0 = 1.03\pm 0.03$ & $\pm 0.001$ \\
%%  & & \\
\hline
2. Density perturbations & & \\
{\ \ \ }a. Adiabatic:  acoustic peaks & at least 3 & 6 or 7 \\
{\ \ \ }b. Gaussian  & no evidence against &  ?? \\
{\ \ \ }c. $ |n-1| \sim {\cal O}(0.1) $ & $n=1.05\pm 0.09$ & $\pm 0.001$  \\
{\ \ \ }d. $dn/d\ln k \sim {\cal O}(10^{-3})$ & $ dn/d\ln k = -0.02\pm 0.04$ & $\pm 10^{-3}$\\
{\ \ \ }e. CDM Paradigm   & many successes & \\
%% & & \\
\hline
3. Gravity waves & &                                  \\
{\ \ \ }a. Amplitude & $T/S < {\cal O}(1)$  & $T/S > 10^{-3}$   \\
{\ \ \ }b. $ n_T\sim {\cal O}(-0.1)$ & -- &  $\pm 0.03$ ?? \\
{\ \ \ }c. Consistency: $T/S = -5 n_T$ & -- & --  \\ \hline
\end{tabular}
%%\end{center}
\end{table}

\section{How much of the truth does inflation have and who is $\phi$?}

A key to testing inflation and getting at the underlying physics is
the detection of gravitational waves.   Not only is the amplitude
of the gravitational waves directly proportional to the energy scale
of inflation, $T\ \propto \ V/\mpl^4$, but this third prediction of 
inflation is an undeniable smoking gun for inflation.

Let me elaborate and comment.  Some would claim that the first two predictions
of inflation -- flat Universe and scale-invariant density
perturbations -- have long been considered features of any sensible
cosmological model.  Certainly both were discussed well before
inflation (e.g., flatness by Peebles and Dicke \cite{dp} and scale-invariant
density perturbations by Harrison and Zel'dovich \cite{hz}).
Thus, there is some truth to this point of view.  However, because
inflation provides a mechanism for actually producing
a flat Universe with almost scale-invariant density perturbations it also makes
a prediction about how close to scale invariant they should be.  As I
have emphasized, an important test of inflation is its prediction
that $|n-1| \sim {\cal O}(0.1)$ and not $n=1$.

In addition to providing a smoking gun for inflation, the detection of gravity
waves instantly reveals the epoch and energy scale of inflation:
\begin{eqnarray*}
H_I^{-1} & \simeq & {2\times 10^{-39}\,{\rm sec}\over \sqrt{T/S}} \\
V^{1/4} & = & 3\times 10^{-3}\mpl\,(T/S)^{1/4} \simeq 3\times 10^{16}\,{\rm GeV}(T/S)^{1/4}
\end{eqnarray*}

Further, the values of $T/S$ and $n-1$ can be used to solve for the
inflationary potential and its first two derivatives:
\begin{eqnarray}
V & = & 1.8 T\, \mpl^4   \\
V^\prime & = & \pm \sqrt{{8\pi \over 5}\,{T\over S}} V/\mpl , \\
V^{\prime\prime} & = & 4\pi \left[ (n-1) + {3\over 5} {T\over S} \right]
V /\mpl^2
\end{eqnarray}
where the numerical factors depend upon the composition
of the Universe and are given for $\Omega_M=0.35$ and $\Omega_\Lambda =0.65$ \cite{TW}.
Measurements of $T$, $S$, and $(n-1)$ can be used to shed light
on the underlying inflaton potential.

\section{If it smells like a rose, is it a rose?}

Inflation has some of the truth, but
it has by no means been verified to the degree that we
can safely include it as part of ``a new standard cosmology.''
Further, while inflation seemed like a bold and expansive
step forward twenty years ago, by today's standards it
seems more modest:   It only explains the isotropy and homogeneity
on a temporary basis (albeit an exponentially long temporary
basis); It does not address the question of the initial singularity (or
why there is a universe at all); and It still stands
disconnected from string theory or any other fundamental
theory of elementary particle physics.  For all of these reasons
we should be open to new ideas.            At the moment
there are several intriguing ideas -- e.g., ekpyrotic/cyclic
model \cite{cyclic} and variable speed of light theories \cite{vsl} --
but none have reached the point of making sharp predictions like inflation.

The cosmological data that we have seen to date have not
only provided the first tests of inflation, but they have
also raised the bar for its alternatives.  In fact, a die-hard
inflationist might well argue that any theory
that is able to account for the observable facts without
appealing to initial conditions will have to look very
much like inflation.

To be specific, the existence of adiabatic
density perturbations on superhorizon scales at the
time of last scattering and the enormous heat of the big
bang (quantified by the entropy of $10^{90}$ within
our current Hubble volume) require any scenario
that does not simply appeal to initial conditions
to incorporate superluminal expansion (accelerated expansion
or accelerated contraction) and entropy production,
the two hallmarks of inflation \cite{htw,liddle}.

The necessity of each is easy to explain.  First,
to causally create a density perturbation in an expanding Universe
requires that a sub-hubble-scale sized region at very early times ($l\la H^{-1}$) must grow
to a size much, much greater than the Hubble length at last scattering.
This translates into a kinematic requirement:  There must be an epoch where the scale factor
grows faster than $t$.  (More precisely and more generally, the condition is that $\dot R$
and $\ddot R$ have the same sign, which is called superluminal expansion.)

Today's Hubble volume contains an entropy of about $10^{90}$ (largely in the
form of photons and neutrinos).  At early times, the entropy within a
Hubble volume is at most $(\mpl /T)^3$ (equality pertaining for
a radiation dominated phase).  In the absence of massive entropy production
the superluminal phase will produce very large, but very empty (low entropy)
density perturbations.  Entropy production is needed to create enough
photons and eventually enough matter to account for the $10^{68}$
baryons (and even more dark matter particles) per galaxy.
(This same argument could have been worded in
terms of producing a smooth region of the Universe corresponding to
our present Hubble volume.)

To summarize, an alternative to inflation must
involve superluminal expansion and massive entropy production; in addition,
if it can be described in 4-dimensions by a the evolution of a
single degree of freedom (which could be called a scalar field), 
an inflation hegemonist might be tempted to call it a realization
of inflation, rather than a fundamentally new paradigm.  She might
have a good case, since inflation is still a paradigm in search of a model.

\section{Concluding remarks}

The chance that any major new idea in theoretical physics has something
to do with the truth is very small; in cosmology the odds are even longer.
Nonetheless, inflation has passed its first round of major tests and
it does appear that inflation has some of the truth.

Precision measurements of CMB anisotropy on sub-degree angular scales
by ground-based and balloon-borne experiments
have led the way in testing inflation.
Over the next decade or so, results from satellite-borne
experiments with even greater precision and control
of systematics -- MAP, Planck
and possibly a new satellite mission dedicated to polarization --
will sharpen the tests by more than 30 fold and the
crucial gravity-wave signature of inflation may be detected.

The challenge and importance of verifying the third prediction of
inflation cannot be overstated.  Discovery of inflation-produced
gravity waves is a smoking gun signature of inflation, immediately identifies
the epoch and scale of inflation and reveals information about the inflationary
potential.  If its spectral index can be measured,
the consistency of the single-field inflation paradigm can be checked.

While it is too early to bring on the champagne -- plenty could still
go wrong -- the bar has been raised for any new competitor to inflation.
Accounting for the superhorizon-sized adiabatic density perturbations
whose existence has been confirmed by detection of acoustic peaks in
the CMB angular power spectrum requires both superluminal expansion
and entropy, the two hallmarks of inflation.

While the successes of inflation are gratifying, we are very far
from understanding the underlying cause.  Single-field models
require very weakly coupled scalar fields (dimensionless couplings
of order $10^{-14}$ or so) and the potentials seem hopelessly contrived.
It could be that our 4-dimensional scalar-field formulation
of inflation makes things look unnatural, and that
when viewed in higher dimensions or when the scalar field is
properly interpreted, the inflaton and its potential will make perfectly
good sense.

In addition to the ``who is $\phi$ question'', there are other 
questions:  the details of reheating, the possibility
of quantum fluctuations during inflation excite isocurvature modes
or produce particle relics, the search for a connection between
planck scale physics and CMB anisotropy (the modes seen on the
CMB sky today were subPlanckian in size during inflation),
and the multiverse.

Perhaps the most daunting challenge facing inflation is convincing
even ourselves that inflation {\em really} happened.  Even if
inflation passes all of its tests and its gravity waves
are detected both by their CMB signature and directly
by Hz-frequency laser interferometers, will we be
able to say, with the same confidence that use in discussing
big-bang nucleosynthesis or the quark/hadron transition, that
the Universe really did inflate (rather than inflation provides
a consistent way of describing what we see today)?  Our
confidence in big-bang nucleosynthesis derives from laboratory-based
nuclear physics and in the quark/hadron transition from computer
simulations and accelerator experiments.  Will we be to find a
laboratory crosscheck for inflation that will gives us similar
confidence?

Finally, there is a downside to inflation:
If inflation does succeed in making the present state of the Universe
insensitive to its initial state, it will place a veil between 
us the beginning of the Universe.

\paragraph{Acknowledgments.}  This work was supported by the DoE (at
Chicago and Fermilab), NSF (at the CfCP) and by the NASA
(at Fermilab by grant NAG 5-7092).

\end{document}